\documentclass[letterpaper, 10 pt, conference]{ieeeconf}
\usepackage{cite}
\usepackage{amsmath,amssymb,amsfonts}
\usepackage{algorithm}
\usepackage{algpseudocode}
\usepackage{graphicx}
\usepackage{textcomp}
\usepackage{tikz}
\usepackage{subcaption}

\usepackage{amsthm}
\newtheorem{theorem}{Theorem}
\newtheorem{lemma}{Lemma}

\newtheorem{corollary}{Corollary}

\newtheorem{definition}{Definition}

\newtheorem{remark}{Remark}
\newtheorem{assumption}{Assumption}

\usepackage{mdframed}

\newcommand{\reals}{\mathbb{R}}
\newcommand{\complex}{\mathbb{C}}
\newcommand{\nats}{\mathbb{N}}
\newcommand{\identity}{\mathbb{I}}

\IEEEoverridecommandlockouts

\AtBeginDocument{%
\setlength{\abovedisplayskip}{5pt plus 2pt minus 3pt}%
\setlength{\belowdisplayskip}{5pt plus 2pt minus 3pt}%
\setlength{\abovedisplayshortskip}{2pt plus 2pt}%
\setlength{\belowdisplayshortskip}{2pt plus 2pt minus 1pt}%
}

\def\BibTeX{{\rm B\kern-.05em{\sc i\kern-.025em b}\kern-.08em
    T\kern-.1667em\lower.7ex\hbox{E}\kern-.125emX}}

\usepackage{eso-pic}
\AddToShipoutPictureFG*{%
  \AtPageUpperLeft{\raisebox{-0.42in}{\makebox[\paperwidth][c]{\normalsize\itshape
    Accepted to the 65th IEEE Conference on Decision and Control (CDC 2026)}}}%
  \AtPageLowerLeft{\raisebox{0.32in}{\makebox[\paperwidth][c]{\parbox{0.84\paperwidth}{\centering\scriptsize
    \copyright\,2026 IEEE. Personal use of this material is permitted. Permission from IEEE must be obtained for all
    other uses, in any current or future media, including reprinting/republishing this material for advertising or
    promotional purposes, creating new collective works, for resale or redistribution to servers or lists, or reuse
    of any copyrighted component of this work in other works.}}}}}
\begin{document}
\title{Safe Autonomous Takeover for Unknown Descriptor Systems\vspace{-0.7em}}
\author{Ben Wooding$^{1}$, Mohamad H. Kazma$^{2}$, Hongchao Zhang$^{1}$, \\ Ahmad F. Taha$^{2}$, Meiyi Ma$^{1}$, and Taylor T. Johnson$^{1}$%
\thanks{$^{1}$ Institute for Convergent Software Integrated Systems, $^{2}$ Department of Civil and Environmental Engineering, Vanderbilt University, Nashville, TN 37235 USA. Corresponding author: \texttt{ben.wooding@vanderbilt.edu}} \thanks{The material presented in this paper is based upon work supported by the National Science Foundation (NSF) through grant numbers 2220401, 2443803, 2152450 and 2550934, and the Defense Advanced Research Projects Agency (DARPA) under contract number FA8750-23-C-0518. Any opinions, findings, and conclusions or recommendations expressed in this paper are those of the authors and do not necessarily reflect the views of DARPA or NSF.}}

\maketitle

\begin{abstract}
Discrete-time descriptor systems are systems with dynamics described by difference equations that are subject to algebraic constraints. These constraints make standard state-space control design approaches significantly more challenging. Recent works have designed control barrier certificates (CBCs) to ensure safety of \emph{unknown} discrete-time systems via data-driven parameterization methods. However, the singular nature of a descriptor system ensures that collected data trajectories will not satisfy the \emph{full-row rank condition} required for persistence of excitation. Essentially, the data will not be rich enough to replace knowing the full descriptor system. In this work, we derive an algorithmic approach to design a safe control policy \emph{online} for unknown discrete-time \emph{nonlinear} descriptor systems, certified by a CBC. We prove that data collected from the subspace described by the algebraic constraints is rich enough to guarantee persistence of excitation. We then demonstrate \emph{autonomous takeover}, a user (who selects safe input choices until autonomous takeover is ready) passes system control fully to a safe autonomous control strategy, guaranteeing safe behavior for all future time steps. We consider two unstable descriptor systems; an ecological case study and a (\emph{trigonometric}) constrained inverted pendulum.
\end{abstract}

\section{Introduction and Paper Contributions}
\label{sec:introduction}

A discrete-time descriptor system is a mathematical model for dynamic systems where the state evolution is defined by a mixture of difference equations and algebraic constraints. Both discrete-time and continuous-time descriptor systems~\cite{belov2018control,dai1989singular,kunkel2024differential,campbell2019applications} describe many constrained real-world applications, and have significant relevance to system-theoretic analysis. Verifying safety properties (preventing unsafe behavior) for such systems is therefore of significant worth. While control barrier certificates (CBCs)~\cite{wieland2007constructive} are an established method to guarantee system safety for difference equations~\cite{agrawal2017discrete} and ordinary differential equations~\cite{xiao2019control}, little research has been conducted to guarantee safety for descriptor systems.

Few works combine formal approaches with descriptor system analysis, with the power system domain primarily considered. The work~\cite{pedersen2016verification} considers explicit differential-algebraic barrier certificate verification but no control design. The work~\cite{kundu2019distributed} synthesizes distributed barrier-based control policies on polynomial approximations of microgrid dynamics. The works~\cite{althoff2013reachability,althoff2014formal} consider reachability analysis to formally verify descriptor power systems. Only the works~\cite{park2024differential} and \cite{zhang2026verification} design a control policy to guarantee safety while fully considering the algebraic constraints. In this setting, our work is unique as it is the first work to design CBCs for \emph{discrete-time} descriptor systems; we also derive all results \emph{from data} assuming \emph{unknown} system coefficients.

Data-driven discrete-time CBC approaches have been explored in~\cite{samari2024singletrajectory,wooding2024learning,wooding2026k,edwards2024fossil,gardner2025trust}, leveraging data parameterization for controller synthesis for non-algebraic systems. The work~\cite{murali2022scenario} uses the scenario approach to design barrier certificates purely from data, but does not consider controller design. We consider that the descriptor systems under analysis have unknown matrix coefficients.

Leveraging results from the data-parameterization literature~\cite{de2019formulas,schmitz2022willems,verhoek2021fundamental}, we demonstrate how a single input-state trajectory can provide safety guarantees for an unknown \emph{nonlinear} descriptor system. The considered conditions build upon the foundation of Willems' fundamental lemma~\cite{willems2005note}, but are more relaxed, with non-consecutive data trajectories being viable. Such results for descriptor systems have been considered by~\cite{schmitz2022willems} and \cite{zhang2025data}. However, these works consider only discrete-time \emph{linear} descriptor systems, while we demonstrate success for \emph{nonlinear} descriptor systems.

\noindent\textbf{Contributions.} This work studies data-driven control of nonlinear descriptor systems with formal guarantees. Our paper contributes the following:
\begin{itemize}
    \item We provide sufficient conditions to design a data-driven control policy and CBC, guaranteeing safety for \emph{unknown discrete-time nonlinear} descriptor systems;
    \item We show theoretically how the persistency of excitation (PE) rank condition can be satisfied by projecting the dynamic equation onto the algebraic subspace during data collection, certifying the data is rich enough to use in place of knowing the true system;
    \item We demonstrate a novel \emph{online} algorithmic framework, enabling a control policy to conduct \emph{autonomous takeover} from a user once safety is assured. We demonstrate the approach on two unstable nonlinear index-$1$ descriptor systems, including a \emph{trigonometric} nonlinear term (without approximation).
\end{itemize} 

\noindent\textbf{Notation.}
 We use $\reals,$ $\reals_{\geq 0}$, $\reals_{>0}$, $\complex$, $\nats$ and $\nats_{>0}$ to denote real numbers, non-negative real numbers, positive real numbers, complex numbers, and non-negative and positive integers, respectively. $\identity_n$ is the $\reals^{n\times n}$ identity matrix. Respectively, $\mathrm{rank}(A)$ and $\mathrm{det}(A)$ denote the row rank and determinant of matrix $A$. $A^\dagger$ indicates the Moore-Penrose pseudoinverse of $A$. We denote the empty set by $\emptyset$.

 \noindent\textbf{Organization.}
 Section~\ref{sec:problem} presents necessary background. Section~\ref{sec:projection} outlines projecting the dynamics of the descriptor system onto its algebraic subspace. Section~\ref{sec:dd-cbc} derives the data-driven approach to guarantee safety. Section~\ref{sec:online} shows our algorithm. Section~\ref{sec:case-studies} demonstrates the approach on two examples. Finally, Section~\ref{sec:conclusion} concludes.

\section{Key Concepts and Problem Statement}
\label{sec:problem}

The descriptor systems under study are nonlinear with unknown matrix coefficients and evolve in discrete-time. Using data, we will design a control policy that guarantees the safety of such a system.

\begin{definition}[Nonlinear Descriptor System]
	\label{def:descriptor}
	A discrete-time nonlinear descriptor system in this work is described by
\begin{equation}
 \label{eq:descriptor}
E\mathbf{x}(k+1) = A\mathfrak{D}(\mathbf{x}(k)) + B\mathbf{u}(k),
\end{equation}
	with consistent initial condition $\mathbf{x}(0)$, where $\mathbf{x}(k)\in\reals^{n}$ are system states, $\mathbf{u}(k)\in\reals^{m}$ are control inputs, and $\mathfrak{D}(\mathbf{x}(k))\in\reals^N$ is a dictionary of \emph{known} nonlinear terms. The matrices $A\in\reals^{n\times N}, E\in\reals^{n\times n},$ and $B\in\reals^{n\times m}$ are all assumed to be \emph{unknown}.
\end{definition}

When clear from context, we ease notation by dropping the variable $k$, \emph{i.e.} we use $\mathbf{x}$ for $\mathbf{x}(k)$, and $\mathbf{x}^+$ for $\mathbf{x}(k+1)$.

\begin{assumption}
\label{ass:separable}
    We assume $\mathfrak{D}(\mathbf{x}) = \Theta(\mathbf{x})\mathbf{x}$ is known, where $\Theta(\mathbf{x})$ is of appropriate dimension.
\end{assumption}

While we assume that the coefficient matrices are \emph{unknown}, reflecting practical scenarios, the dictionary $\mathfrak{D}$ is assumed to be available. The dictionary should be \emph{sufficiently comprehensive} to capture the true nonlinearities in the actual system even if irrelevant terms are included. This decomposition is widely used~\cite{chen2001atomic}, \emph{e.g.,} the dictionary $\mathfrak{D}$ may be derived from first principles, despite the specific coefficients remaining unknown. We make a semantic choice to describe this system as \emph{unknown}. If matrix $E$ in~\eqref{eq:descriptor} is nonsingular, then the system can be rewritten as an explicit difference equation, \[\mathbf{x}^+ = E^{-1}A\mathfrak{D}(\mathbf{x}) + E^{-1}B\mathbf{u}.\]

This paper therefore focuses on the descriptor system~\eqref{eq:descriptor} when $E$ is singular, $\mathrm{rank}(E) < n$,  and the system is \emph{causal}.

\subsection{Generating Persistently Excited Data}
\label{sec:data-drivenmodels}

Data parameterization and persistency of excitation (PE) has been used to \emph{exactly} represent the behavior of an unknown system for controller design using data. We represent the state-input data collected from the unknown descriptor over the time horizon $[0,T]$, where $T\in\mathbb{N}_{>0}$, as:
\begin{subequations}
\label{eq:data}
\begin{align}
	\label{eq:state-data}
	&\mathbb{X}_{0} = [~\mathbf{x}(0),~\mathbf{x}(1),~\ldots,~\mathbf{x}(T-1)] \in\reals^{n\times T},\\
    \label{eq:derivative-data}
	&\mathbb{X}_{1} = [\mathbf{x}(1),\mathbf{x}(2),\ldots,\mathbf{x}(T)] \in\reals^{n\times T},\\
    \label{eq:input-data}
	&\mathbb{U}_0 = [\mathbf{u}(0),\mathbf{u}(1),\ldots,\mathbf{u}(T-1)] \in\reals^{m\times T}.
\end{align}
We also include the data representation of the extensive dictionary of nonlinear functions:
\begin{align}
	\mathbb{D}_0 = [\mathfrak{D}(\mathbf{x}(0)),\ldots,\mathfrak{D}(\mathbf{x}(T-1))], \in\reals^{N\times T}.
\end{align}
\end{subequations}

We name~\eqref{eq:data} collectively as a single input-state trajectory. To ensure PE, the collected data must satisfy full-row rank conditions of the data matrices~\eqref{eq:data}, this fact is related to the famous \emph{fundamental lemma}~\cite{willems2005note}. For model identification~\cite{de2019formulas}, it is necessary that, $\mathbb{D}_0$ and $\mathbb{U}_0$ are full-row rank, \[\mathrm{rank}\begin{bmatrix}
    \mathbb U_0\\ \mathbb{D}_{0}
\end{bmatrix} = N + m.\] 
For the simpler case of learning a control policy~\cite{de2023cancellation}, it is sufficient that only $\mathbb{D}_0$ is full-row rank, \begin{equation*}
    \mathrm{rank}\begin{bmatrix}
        \mathbb{D}_{0}
    \end{bmatrix} = N,
\end{equation*}
to guarantee PE. For the derivations of these conditions, we refer to the original works~\cite{de2019formulas,de2023cancellation}. Crucially, applying this approach directly to~\eqref{eq:descriptor} fails. With $E$ singular, the matrices cannot satisfy the full-row rank conditions required to parameterized the descriptor system.

\subsection{Control Barrier Certificates}
\label{sec:CBC}

In this study we design a control policy such that all trajectories evolving from an initial region remain safe. The following definition defines safety.

\begin{definition}[Safety]
    The descriptor system~\eqref{eq:descriptor} is safe if all state trajectories evolving from the initial region $\mathbf{x}(0)\in \mathcal{X}_\mathcal{I}$ never reach the unsafe region $\mathcal{X}_\mathcal{U}$, \emph{i.e.}, $\mathbf{x}(k)\notin \mathcal{X}_{\mathcal U},~\forall k\in\nats$.
\end{definition}
A well-established safety guarantee is the synthesis of a control barrier certificate (CBC)~\cite{wieland2007constructive}.
\begin{definition}[CBC]
\label{def:CBC}
    Let $\mathbf{x}^+ = f(\mathbf{x},\mathbf{u})$ be a general discrete-time system and $\mathcal{X}, \mathcal{X}_\mathcal{I}, \mathcal{X}_\mathcal{U}$ the given state space, initial region, and unsafe region, respectively, where $\mathcal{X}_\mathcal{I}, \mathcal{X}_\mathcal{U} \subset \mathcal{X}$ and $\mathcal{X}_\mathcal{I}\cap \mathcal{X}_\mathcal{U}=\emptyset$. Suppose there exists a control policy $\mathbf{u}$ with control barrier certificate $\mathfrak{B} : \mathcal{X} \rightarrow \reals$ that satisfies the following conditions:
\begin{subequations}
\label{eq:CBC-all}
    \begin{align}
    \label{eq:CBC1}
    \mathfrak{B}(\mathbf{x}) &\leq \beta_1, &\forall \mathbf{x}&\in \mathcal{X}_\mathcal{I}\\
    \label{eq:CBC3}
    \mathfrak{B}(f(\mathbf{x},\mathbf{u})) &\leq \mathfrak{B}(\mathbf{x}), &\forall \mathbf{x}&\in \mathcal{X}_\mathcal{I}\\
     \label{eq:CBC2}
    \mathfrak{B}(\mathbf{x}) &\geq \beta_2, &\forall \mathbf{x}&\in \mathcal{X}_\mathcal{U}
\end{align}
\end{subequations}
where $\beta_1,\beta_2\in\reals$ are the sublevel sets of $\mathfrak{B}(\mathbf{x})$, and $\beta_2 > \beta_1$, then the safety of the discrete-time system is guaranteed.
\end{definition}

These safety conditions are analogous to Lyapunov functions. Since, the unsafe region $\mathcal{X}_\mathcal{U}$ maps to larger values than the values of the initial region $\mathcal{X}_\mathcal{I}$, then, a control policy $\mathbf{u}$ ensuring the barrier value of any state transitions (initialized in $\mathcal{X}_\mathcal{I}$) are non-increasing assures safety.

\subsection{Problem Statement}

We now describe the problem we address in this paper. 
\begin{mdframed}
\textbf{Problem Statement.}
Consider the singular system~\eqref{eq:descriptor} with unknown matrices
$E$, $A$ and $B$, a known dictionary $\mathfrak{D}(\mathbf{x}(k))$, and initial and unsafe sets $\mathcal{X}_{\mathcal{I}}, \mathcal{X}_{\mathcal{U}} \subset \mathcal{X}$ where $\mathcal{X}_{\mathcal{I}} \cap \mathcal{X}_{\mathcal{U}} = \emptyset$. Design a safe control policy $\mathbf{u}(k)$ certified by a CBC $\mathfrak{B}(\mathbf{x}(k))$, using a single input-state trajectory~\eqref{eq:data} collected from the user-controlled unknown system, and enabling the online transfer of system control from user to the safe autonomous policy.
\end{mdframed}

In previous work, the data-driven control policy design for ODE systems has been hierarchical (\emph{e.g.}~\cite{gardner2025trust}): (i) a persistently excited data trajectory is collected, (ii) a CBC and safe control policy are designed offline, (iii) the controller is deployed to the system. We present here the novelty of collecting the data and designing the control policy online, our algorithmic approach can alert a user when \emph{autonomous takeover} is possible.

\section{Projection to Algebraic Subspace}
\label{sec:projection}

 We motivate this section using linear systems.

\subsection{Discrete-Time Linear Descriptor Systems}

\begin{definition}[Linear Descriptor System] A discrete-time linear descriptor system is described by
	\label{def:lin-descriptor}
\begin{equation}
 \label{eq:lin-descriptor}
E\mathbf{x}(k+1) = A\mathbf{x}(k) + B\mathbf{u}(k),
\end{equation}
	with consistent initial condition $\mathbf{x}(0)$, where $\mathbf{x}(k)\in\reals^{n}$ are system states and $\mathbf{u}(k)\in\reals^{m}$ are control inputs. The matrices $A,E\in\reals^{n\times n}, $ and $B\in\reals^{n\times m}$.
\end{definition}

\begin{assumption}[Regularity]
    We make a standard assumption throughout this work that $\mathrm{det}(\lambda E - A) \neq 0 $ for some $\lambda\in\complex$, \emph{i.e.} system~\eqref{eq:lin-descriptor} is regular.
\end{assumption}

Regularity is standard in practice, \emph{e.g.} for power systems~\cite{gross2016solvability}. Ensuring the well-posedness and uniqueness of descriptor systems under analysis. Since~\eqref{eq:lin-descriptor} is regular, the following lemma describes a transformation to a semi-explicit representation, enabling the separation of difference equations and algebraic constraints. Computing this form is described further in~\cite[Chap.~8]{dai1989singular}.

\begin{lemma}[Canonical Form]
\label{lem:canonical}
Suppose~\eqref{eq:lin-descriptor} is regular. There exist nonsingular matrices $S\in\reals^{n\times n}$ and $Q\in\reals^{n\times n}$ such that:
\begin{equation*}
    SEQ = \begin{bmatrix}
        \identity_{n_1} & 0\\
        0 & N
    \end{bmatrix}, \quad SAQ = \begin{bmatrix}
        A_1 & 0\\
        0 & \identity_{n_2}
    \end{bmatrix}, \quad SB = \begin{bmatrix}
        B_1\\
        B_2
    \end{bmatrix}, 
\end{equation*}
where $A_1\in\reals^{n_1\times n_1}$ and $N\in\reals^{n_2\times n_2}$ is nilpotent, with nilpotency index $\mu\in\nats$ and $n_1 + n_2 = n$. With a change of variables, \[\mathbf{x} = Q\begin{bmatrix}
    \mathbf{x}_1 \\ \mathbf{x}_2
\end{bmatrix}.\] 
The canonical form is then,
\begin{equation*}
    \begin{bmatrix}
        \identity_{n_1} & 0\\
        0 & N
    \end{bmatrix}\begin{bmatrix}
    \mathbf{x}_1^+ \\ \mathbf{x}_2^+
\end{bmatrix} =\begin{bmatrix}
        A_1 & 0\\
        0 & \identity_{n_2}
    \end{bmatrix} \begin{bmatrix}
    \mathbf{x}_1 \\ \mathbf{x}_2
\end{bmatrix} + \begin{bmatrix}
        B_1\\
        B_2
    \end{bmatrix}\mathbf{u}.
\end{equation*}
\end{lemma}

We therefore acquire two equations, a forward recurrent equation uniquely determined by $\mathbf{x}(0)$ and $\mathbf{u}(k)$,
\begin{equation*}
    \mathbf{x}_1(k) = A_1^k\mathbf{x}_1(0) + \sum_{i=0}^{k-1}A_1^{k-i-1}B_1\mathbf{u}(i),
\end{equation*}
and a backward recurrent equation, described using the terminal time instance $L$, and uniquely determined by the terminal state $\mathbf{x}_2(L)$ and $\mathbf{u}(k)$,
\begin{equation}
    \label{eq:backward-recurrent}
    \mathbf{x}_2(k) = N^{L-k}\mathbf{x}_2(L) + \sum_{i=0}^{L-k-1}N^iB_2\mathbf{u}(k+i).
\end{equation}

Therefore the state at any time point in a descriptor system is related to the initial state $\mathbf{x}(0)$ and former inputs, as well as the terminal state $\mathbf{x}(L)$ and future inputs up to time step $L$. When the descriptor system can be described purely by its initial state $\mathbf{x}(0)$ and former inputs $\mathbf{u}(i)$ for $i=0,1,\ldots k-1$, the system is \emph{causal}. From~\eqref{eq:backward-recurrent}, it is clear when $N\neq 0$, then the system is \emph{non-causal}, and the descriptor system may be described by future unseen inputs $\mathbf{u}(k+i)$.

\begin{definition}[Nilpotency Index~\cite{dai1989singular}]
    The smallest $\mu\in\nats$, such that $N^\mu=0$. 
\end{definition}

If the nilpotency index is known, then~\eqref{eq:backward-recurrent} simplifies to the substate
\begin{equation*}
    \mathbf{x}_2(k) = \sum_{i=0}^{\mu-1}N^iB_2\mathbf{u}(k+i),
\end{equation*}
however, future inputs would still be required, and the formulation may be non-causal. System~\eqref{eq:lin-descriptor} is causal when the \emph{nilpotency index} $\mu \in \{0,1\}$.

In our problem setting, the nonlinear system~\eqref{eq:descriptor} has constant matrices $E,A,B$ (only $\mathfrak{D}(\mathbf{x})$ is nonlinear), so the pencil $(\lambda E-A)$ is still well-defined and an equivalent canonical form exists for~\eqref{eq:descriptor} (admitting a semi-explicit representation). We therefore directly continue with the regularity assumption and nilpotency index conceptualization in our nonlinear setting. As we consider only causal systems, and with $\mu =0$ a trivial case (non-singular), we limit our study to nonlinear descriptor systems with nilpotency index $\mu = 1$.

\subsection{Projection of Nonlinear System}

To represent~\eqref{eq:descriptor} using parameterized data, we need to satisfy the before mentioned full-row rank conditions. The chief culprit of this failure is the algebraic constraints caused by matrix $E$ being singular. By Lemma~\ref{lem:canonical}, consider a \emph{causal} semi-explicit representation of~\eqref{eq:descriptor}:
\begin{equation*}
    \overbrace{\begin{bmatrix} \identity_{n_d} & 0 \\ 0 & 0 \end{bmatrix}}^{E}
  \overbrace{\begin{bmatrix} \mathbf{x}_d^+ \\ \mathbf{x}_a^+ \end{bmatrix}}^{\mathbf{x}^+}
  =
  \overbrace{\begin{bmatrix} A_{11} & A_{12} \\ A_{21} & A_{22} \end{bmatrix}}^{A}
  \overbrace{\begin{bmatrix} \mathfrak{D}_d(\mathbf{x}_d) \\ \mathfrak{D}_a(\mathbf{x}_a) \end{bmatrix}}^{\mathfrak{D}(\mathbf{x})}
  +
  \overbrace{\begin{bmatrix} B_1 \\ B_2 \end{bmatrix}}^{B}
  \mathbf{u},
\end{equation*}
where $\mathbf{x}_d(k)\in\reals^{n_d}$ and $\mathbf{x}_a(k)\in\reals^{n_a}$ are the separated dynamic states and algebraic states of the system $\mathbf{x} = [\mathbf{x}_d^\top~\mathbf{x}_a^\top]^\top$, and $n_d+n_a = n$. Similarly, we extend Assumption~\ref{ass:separable} to partition the dictionary $\mathfrak{D}(\mathbf{x})$ into $\mathfrak{D}_d(\mathbf{x}_d)\in\reals^{N_d}$ and $\mathfrak{D}_a(\mathbf{x}_a)\in\reals^{N_a}$, where $\mathfrak{D}_i(\mathbf{x}_i) = \Theta_i(\mathbf{x}_i)\mathbf{x}_i$ for $i\in\{d,a\}$, and $N_d + N_a = N$. The introduced sub-matrices are considered to be of appropriate size, and $\mathbf{u}$ is unchanged. We simplify to get the following:
\begin{subequations}
\label{eq:semi-explicit-descriptor}
\begin{align}
    \label{eq:semi-explicit-x1}
    \mathbf{x}_d^+ &=  A_{11}\mathfrak{D}_d(\mathbf{x}_d) +A_{12}\mathfrak{D}_a(\mathbf{x}_a) + B_1\mathbf{u},\\
    \label{eq:semi-explicit-x2}
    0 &= A_{21}\mathfrak{D}_d(\mathbf{x}_d) +A_{22}\mathfrak{D}_a(\mathbf{x}_a)+ B_2\mathbf{u},
\end{align}
\end{subequations}

To satisfy the rank condition, we intend to project the dynamics of the system~\eqref{eq:semi-explicit-x1} onto the algebraic subspace $\mathcal{S} = \{\mathbf{x}\in\reals^n : A_{21}\mathfrak{D}_d(\mathbf{x}_d) + A_{22}\mathfrak{D}_a(\mathbf{x}_a) + B_2\mathbf{u} = 0\}$, described by~\eqref{eq:semi-explicit-x2}. This projection lets us consider a reduced-order difference equation that satisfies the rank condition while not abstracting away any crucial impulse behaviors that can destroy the system in practice~\cite{dai1989singular}.

We present the following theorem, providing the reduced-order difference equation that is a projection of~\eqref{eq:semi-explicit-x1} onto the constraints subspace~\eqref{eq:semi-explicit-x2}. We call this new system description the \emph{projected system} or \emph{projected model}.

\begin{theorem}[Projected System]
\label{thm:projected}
For nonsingular matrix $A_{22}$, the dynamics $\mathbf{x}_d$ in the causal full descriptor system~\eqref{eq:semi-explicit-descriptor}, when the trajectories lie on the algebraic subspace~\eqref{eq:semi-explicit-x2}, are exactly equivalent to the difference equation
\begin{equation}
    \label{eq:descriptor-x1}
    \mathbf{z}^+ = \Tilde{A}\mathfrak{D}_d(\mathbf{z}) + \Tilde{B}\mathbf{u}, 
\end{equation}
where $\mathbf{z} = \mathbf{x}_d$, $\Tilde{A} = A_{11} - A_{12}A_{22}^{-1}A_{21},$ and
        $\Tilde{B} = B_1-A_{12}A_{22}^{-1}B_2$.
\end{theorem}
\begin{proof}
    With $A_{22}$ nonsingular, we can rewrite~\eqref{eq:semi-explicit-x2} as the following:
    \begin{equation*}
        \mathfrak{D}_a(\mathbf{x}_a) = -A_{22}^{-1}[A_{21}\mathfrak{D}_d(\mathbf{x}_d) + B_2\mathbf{u}].
    \end{equation*}
    We project the dynamics into the subspace by canceling $\mathfrak{D}_a(\mathbf{x}_a)$ in~\eqref{eq:semi-explicit-x1}:
    \begin{align*}
        \mathbf{x}_d^+ = [A_{11} -A_{12}A_{22}^{-1}A_{21}]\mathfrak{D}_d(\mathbf{x}_d) + [B_1-A_{12}A_{22}^{-1}B_2]\mathbf{u}.
    \end{align*}
    Substituting  $\mathbf{z} = \mathbf{x}_d$, $\Tilde{A} = A_{11} - A_{12}A_{22}^{-1}A_{21},$ and
        $\Tilde{B} = B_1-A_{12}A_{22}^{-1}B_2$ completes the proof.
\end{proof}

The matrices $\Tilde{A}$ and $\Tilde{B}$ abstract away the unknown matrices of~\eqref{eq:semi-explicit-descriptor}, and are themselves unknown. Using the parameterization approach of this work we never need to compute $\Tilde{A}$ and $\Tilde{B}$ explicitly, and they are eliminated when constructing the conditions that require only the single input-state trajectory.

\begin{remark}
 As $\mathcal{S}\subset \mathcal{X}$, the initial and unsafe regions on the subspace can be redefined as $\mathcal{X}_{\mathcal{I}} := \mathcal{S}\cap \mathcal{X}_{\mathcal{I}}$ and $\mathcal{X}_{\mathcal{U}} := \mathcal{S}\cap \mathcal{X}_{\mathcal{U}}$, respectively.
\end{remark}

\section{Data-Driven Control Barrier Certificates}
\label{sec:dd-cbc}

The previous section demonstrated how we can consider the projected system~\eqref{eq:descriptor-x1} in place of the descriptor system~\eqref{eq:semi-explicit-descriptor} by projecting the dynamics onto the algebraic constraints. To design a control policy and synthesize a CBC, we can collect a single input-state trajectory relating to the projected system~\eqref{eq:descriptor-x1} instead of the descriptor system~\eqref{eq:semi-explicit-descriptor}, as the latter does not satisfy the full-row rank condition required for PE. To collect the correct trajectory, we therefore require the following assumptions:

\begin{assumption}[Subspace Data Collection]
\label{ass:subspace-data}
The single input-state trajectory satisfies the following:
\begin{enumerate}
    \item the initial condition $\mathbf{z}(0)\in\mathcal{S}$ (\emph{causal solvability});
    \item the trajectory remains in the subspace $\mathcal{S}$;
    \item the system is \emph{impulse-free},  (system is causal and independent of future inputs).
\end{enumerate}
\end{assumption}

These assumptions are not unduly restrictive. The constraints subspace~\eqref{eq:semi-explicit-x2} relates to real phenomena restricting the physical system, \emph{e.g.} Kirchhoff's laws or robot joint overextension, of which the user can have some prior understanding. Assumption~\ref{ass:separable} is justified by deriving the dictionary from first principles, where spurious additional terms are tolerated. The user has total freedom in how to collect the data trajectory, choosing a consistent initial condition and any input that keeps the trajectory in $\mathcal{S}$. The user can also enforce safety during data collection by selecting known safe input choices. In practice, the ``user'' need not be human: any pre-existing (uncertified) controller, \emph{e.g.} a legacy stabilizing controller, can generate the initial safe inputs. This is pertinent for fast unstable systems, where manual safe operation is difficult.

We (re)define our collected single trajectory for data collected on the subspace~\eqref{eq:semi-explicit-x2} as follows ($\mathbb{U}_0$ unchanged):
\begin{subequations}
\label{eq:z-data}
\begin{align}
	\label{eq:z-state-data}
	&\mathbb{Z}_{0} = [\mathbf{z}(0),\ldots,\mathbf{z}(T-1)] \in\reals^{n_d\times T},\\
    \label{eq:z-dictionary-data}
    &\mathbb{D}_{0} = [\mathfrak{D}_d(\mathbf{z}(0)),\ldots,\mathfrak{D}_d(\mathbf{z}(T-1))] \in\reals^{N_d\times T},\\
    \label{eq:z-output-data}
	&\mathbb{Z}_{1} = [\mathbf{z}(1),\ldots,\mathbf{z}(T)] \in\reals^{n_d\times T},\\
    \label{eq:z-input-data}
	&\mathbb{U}_0 = [\mathbf{u}(0),\ldots,\mathbf{u}(T-1)] \in\reals^{m\times T}.
\end{align}
\end{subequations}
From~\eqref{eq:descriptor-x1} it is trivial that $\mathbb{Z}_{1} = \Tilde{A}\mathbb{D}_{0} + \Tilde{B}\mathbb{U}_{0}$.

We will derive the conditions for a data-driven control policy $\mathbf{u} = K(\mathbf{z})\mathbf{z}$, that guarantees safety via a CBC. The condition $\mathrm{rank}(\mathbb{D}_0) = N_d$ is sufficient for the data to satisfy PE and the row rank is simple to calculate. We consider the data parameterization approach of~\cite{de2019formulas} which is more relaxed than~\cite{willems2005note}. Notably, the collected data could be non-consecutive, but $T
\geq N_d$ is a necessary condition for PE.

\subsection{Data-Driven Projected Model}

Using the single input-state trajectory~\eqref{eq:z-data}, we derive a closed-form data-driven model of~\eqref{eq:descriptor-x1} with control policy $\mathbf{u}=K(\mathbf{z})\mathbf{z}= \mathbb{U}_0Q(\mathbf{z})\mathbf{z}$.

\begin{theorem}[Data-Driven Model]
\label{thm:dd-model}
Let $Q(\mathbf{z})\in\reals^{T\times n_d}$ be a matrix-valued function such that
\begin{equation}\label{eq:Q-matrix}
\Theta_d(\mathbf{z}) = \mathbb{D}_{0}Q(\mathbf{z}),
\end{equation}
where $\Theta_d(\mathbf{z})\in\reals^{N_d\times n_d}$,
$\mathfrak{D}_d(\mathbf{z}) = \Theta_d(\mathbf{z})\mathbf{z}$ and $\mathbb{D}_{0}$ is full row-rank.
If one synthesizes $\mathbf{u}=K(\mathbf{z})\mathbf{z}= \mathbb{U}_0Q(\mathbf{z})\mathbf{z}$, then the closed-loop system $\mathbf{z}^+=\Tilde{A}\mathfrak{D}_d(\mathbf{z})+\Tilde{B}\mathbf{u}$ has the following data-based representation:
\begin{equation}
\label{eq:dd-model}
\mathbf{z}^+ = \mathbb{Z}_{1}Q(\mathbf{z})\mathbf{z},
\end{equation}
equivalently, $\Tilde{A}\Theta_d(\mathbf{z})+\Tilde{B}K(\mathbf{z}) = \mathbb{Z}_{1}Q(\mathbf{z}).$
\end{theorem}
\begin{proof} $\mathbf{z}^+ = \Tilde{A}\mathfrak{D}_d(\mathbf{z}) + \Tilde{B}\mathbf{u}$ is a projected model exactly representing the descriptor~\eqref{eq:semi-explicit-descriptor} (using Theorem~\ref{thm:projected}).
    Since $\mathbf{u}=K(\mathbf{z})\mathbf{z}= \mathbb{U}_0Q(\mathbf{z})\mathbf{z}$, the closed-loop system~\eqref{eq:descriptor-x1} can be written as
\begin{align*}
   \mathbf{z}^+ & = (\Tilde{A}\Theta_d(\mathbf{z})\!+\!\Tilde{B}K(\mathbf{z}))\mathbf{z} = \begin{bmatrix}
        \Tilde{B} & \Tilde{A}
    \end{bmatrix} \!\!\begin{bmatrix}
        K(\mathbf{z}) \\ \Theta_d(\mathbf{z})
    \end{bmatrix} \mathbf{z}. \end{align*}
Leveraging~\eqref{eq:Q-matrix},
    $
   \mathbf{z}^+ =
    \begin{bmatrix}
        \Tilde{B} & \Tilde{A}
    \end{bmatrix} \!\!\begin{bmatrix}
        \mathbb{U}_0 \\ \mathbb{D}_{0}
    \end{bmatrix} Q(\mathbf{z})\mathbf{z}.
$
Therefore, as $\mathbb{Z}_{1} = \Tilde{A}\mathbb{D}_{0} \!+\! \Tilde{B} \mathbb{U}_0$, then, $\mathbf{z}^+ = \mathbb{Z}_{1}Q(\mathbf{z})\mathbf{z}$, equivalently, $\Tilde{A}\Theta_d(\mathbf{z})+\Tilde{B}K(\mathbf{z}) = \mathbb{Z}_{1}Q(\mathbf{z})$, is the data-based representation of the closed loop~\eqref{eq:descriptor-x1}, completing the proof.
\end{proof}

Inaccuracy in the choice of $\mathfrak{D}(\mathbf{x})$ may arise in two ways: \emph{(i)} $\mathfrak{D}(\mathbf{x})$ includes all true basis functions but also spurious ones (our setting), or \emph{(ii)} $\mathfrak{D}(\mathbf{x})$ omits genuine terms. In the first, $\mathfrak{D}(\mathbf{x})$ remains sufficiently rich, although possibly overparameterized, which \emph{only} increases computational cost (more data and decision variables), but the framework remains valid under the inaccuracy. In the second, $\mathfrak{D}(\mathbf{x})$ lacks true dynamic terms, Theorem~\ref{thm:dd-model} implies that the data-based model no longer represents the true closed-loop system, making the CBC invalid for the real dynamics. Related studies~\cite{de2023cancellation} have considered this situation of neglected nonlinearities. Limited guarantees can be maintained, though controller performance typically degrades and simulations must be confined near the origin.

\subsection{Safety Guarantees via Control Barrier Certificates}

One can guarantee the control policy $\mathbf{u}=K(\mathbf{z})\mathbf{z}$ ensures the true descriptor~\eqref{eq:semi-explicit-descriptor} is always safe by certifying a CBC if the conditions of the following theorem are satisfiable. The PE condition is essential to guarantee the Moore-Penrose pseudoinverse of $\mathbb{D}_0$ exists.

\begin{theorem}[Data-Driven CBC for Projected System]
\label{thm:data-CBC}
Consider the causal descriptor~\eqref{eq:semi-explicit-descriptor} with unknown matrices $A_{11},A_{12},A_{21},A_{22}$, $B_{1}$, and $B_2$, and its data-driven representation~\eqref{eq:dd-model}.
Let $\mathcal{X}_{\mathcal{I}}, \mathcal{X}_{\mathcal{U}} \subset \mathcal{X}$ represent the initial and unsafe regions of the descriptor system, respectively, where $\mathcal{X}_{\mathcal{I}}\cap \mathcal{X}_{\mathcal{U}}=\emptyset$. Suppose there exist constants $\beta_1,\beta_2 \in \reals_{>0}$, where $\beta_2 > \beta_1$, positive-definite matrix $P\in\reals^{n_d\times n_d}$, and $M(\mathbf{z}) = \mathbb{Z}_1\mathbb{D}_0^\dagger\Theta_d(\mathbf{z})$ satisfying the following constraints:
\begin{subequations}
\begin{align}
\label{eq:ct_CBC_data1}
\begin{bmatrix}
    \beta_1 & \mathbf{z}^\top\\
   \mathbf{z} & P^{-1}
\end{bmatrix} &\succeq 0, &\forall \mathbf{z}\in \mathcal{X}_\mathcal{I} \\
\label{eq:ct_CBC_data3}
\begin{bmatrix}
    P^{-1} & M(\mathbf{z})P^{-1}\\
   P^{-1} M(\mathbf{z})^\top & P^{-1}
\end{bmatrix} &\succeq 0, &\forall \mathbf{z}\in \mathcal{X}_\mathcal{I}\\
\label{eq:ct_CBC_data2}
\mathbf{z}^\top P\mathbf{z} &\geq \beta_2, &\forall \mathbf{z}\in \mathcal{X}_{\mathcal{U}}
\end{align}
\end{subequations}
Then $\mathfrak{B}(\mathbf{z}) = \mathbf{z}^\top P\mathbf{z}$ is a CBC and $\mathbf{u} = \mathbb{U}_0\mathbb{D}_{0}^\dagger\mathfrak{D}_d(\mathbf{z})$ is its corresponding safety controller for the collected PE trajectory of the descriptor system~\eqref{eq:semi-explicit-descriptor}.
\end{theorem}

\begin{proof}
Consider Definition~\ref{def:CBC} with quadratic barrier certificate $\mathfrak{B}(\mathbf{x}) = \mathbf{x}^\top P\mathbf{x}$.
Replacing the left-hand side of~\eqref{eq:CBC1} and~\eqref{eq:CBC2} with the above gives:
\begin{align}
    \label{eq:prf-tmp1}
    \mathbf{x}^\top P\mathbf{x} &\leq \beta_1, \quad\quad\quad\forall \mathbf{x}\in \mathcal{X}_{\mathcal{I}}, \\
\mathbf{x}^\top P\mathbf{x} &\geq \beta_2, \quad\quad\quad\forall \mathbf{x}\in \mathcal{X}_{\mathcal{U}}. \nonumber
\end{align}
From Theorem~\ref{thm:projected} and Theorem~\ref{thm:dd-model}, we can directly replace the semi-explicit descriptor system~\eqref{eq:semi-explicit-descriptor} by the data-driven reduced-order model~\eqref{eq:dd-model}. Therefore substitute $\mathbf{x}$ with $\mathbf{z}$ from~\eqref{eq:dd-model}, where $\mathbf{z}= \mathbf{x}_d$. Taking the Schur complement of~\eqref{eq:prf-tmp1} gives~\eqref{eq:ct_CBC_data1}.

Condition~\eqref{eq:ct_CBC_data3} maps to~\eqref{eq:CBC3}, enforcing the barrier function $\mathfrak{B}(\cdot)$ to be non-increasing. It is equivalent to the Lyapunov stability condition for~\eqref{eq:dd-model}:
\begin{align*}
    (\mathbf{z}^+)^\top P\mathbf{z}^+ \leq \mathbf{z}^\top P\mathbf{z}&,\\
    P^{-1} - P^{-1}Q(\mathbf{z})^\top\mathbb{Z}_1^\top P\mathbb{Z}_1Q(\mathbf{z})P^{-1} &\succeq 0,\\
    \begin{bmatrix}
    P^{-1} & \mathbb{Z}_1Q(\mathbf{z})P^{-1}\\
   P^{-1} (\mathbb{Z}_1Q(\mathbf{z}))^\top & P^{-1}
\end{bmatrix} &\succeq 0,
\end{align*}
as $Q(\mathbf{z}) = \mathbb{D}_0^\dagger\Theta_d(\mathbf{z})$, substituting in $M(\mathbf{z})$ completes the proof.
\end{proof}

The results can also be generalized beyond quadratic CBCs to the conditions in Definition~\ref{def:CBC} replacing $\mathbf{x}$ with $\mathbf{z}$ and $f(\mathbf{z,\mathbf{u}})$ with the data-driven projected system description~\eqref{eq:dd-model}. For fixed convex sets describing $\mathbf{z}$,~\eqref{eq:ct_CBC_data1} and~\eqref{eq:ct_CBC_data3} can be solved as LMIs directly as the conditions become linear in $P^{-1}$, and the sublevel set $\beta_1$ is a decision variable to increase the likelihood of valid CBCs. This formulation enables us to move beyond polynomial-type function restrictions, as demonstrated in our second case study. Although the controller design appears decoupled from the CBC, the user has complete freedom in designing the control policy as the user's selection of inputs for $\mathbb{U}_0$ directly affects $\mathbf{u}(k)$. We elaborate on the non-restrictive nature of these conditions in Section~\ref{sec:online}. In the next subsection, we present an alternate control design that is coupled and applied to stability but applicable for safety.

\subsection{Stability Guarantees via Control Lyapunov Functions}

Implicitly we are also studying the design of a control policy for system stability as the CBC condition~\eqref{eq:CBC3} is equivalent to the Lyapunov stability condition of a discrete-time control Lyapunov function (CLF).

\begin{corollary}[Data-Driven CLF]
\label{cor:CLF}
    Consider the unknown descriptor in \eqref{eq:semi-explicit-descriptor}, with its reduced-order projected system's data-driven representation~\eqref{eq:dd-model}. Suppose there exists matrix $H(\mathbf{z})\in\reals^{T\times n_d}$ and positive-definite matrix $P\in\reals^{n_d\times n_d}$, such that $Q(\mathbf{z})=H(\mathbf{z})P$, and constraints,
    \begin{subequations}
    \label{eq:corollary-conditions}
    \begin{align}
\mathbb{D}_0H(\mathbf{z})=\Theta_d(\mathbf{z})P^{-1}\\
\begin{bmatrix}
    P^{-1} & \mathbb{Z}_1H(\mathbf{z})\\
   H(\mathbf{z})^\top\mathbb{Z}_1^\top & P^{-1}
\end{bmatrix} \succeq 0,
    \end{align}
    \end{subequations} 
    are satisfied. Then $\mathfrak{V}(\mathbf{z}) = \mathbf{z}^\top P\mathbf{z}$ is a control Lyapunov function (CLF) and $\mathbf{u} = \mathbb{U}_0H(\mathbf{z})P\mathbf{z}$ is its corresponding stabilizing controller for the unknown descriptor~\eqref{eq:semi-explicit-descriptor}.
\end{corollary}

\section{Online Algorithmic Approach}
\label{sec:online}

We will now use these theoretical guarantees algorithmically. We present a novel controller design strategy that signals online when it is safe to transition over from user inputs to autonomous inputs. The user begins fully controlling the system, then, when the \emph{autonomous takeover} signal is given, the user transfers control to the autonomous control policy that guarantees safety from that instance onward. Our algorithm demonstrates this for the control policy in Theorem~\ref{thm:data-CBC}, but it is also possible to substitute~\eqref{eq:corollary-conditions} for~\eqref{eq:ct_CBC_data3} for the alternative control policy, we mark the differences for the latter using algorithm comments in blue. The online approach is provided in Algorithm~\ref{alg:descriptor}.

We assume we have access to collect data from the descriptor system~\eqref{eq:semi-explicit-descriptor}, and a maximal time horizon $T_{\max}$ is decided in advance that manually terminates the loop. We also assume the initial region $\mathcal{X}_\mathcal{I}$ and unsafe region $\mathcal{X}_\mathcal{U}$ are known, along with the extensive dictionary $\mathfrak{D}_d(\cdot)$. As a useful feature, if no safe controller has been found after several time steps, one can \emph{redefine} $\mathcal{X}_\mathcal{I}$ and $\mathcal{X}_\mathcal{U}$ online at any time step. However, once a safe controller is found these regions can no longer be amended without redesigning the control policy.

\begin{algorithm}[t]
	\begin{algorithmic}[1]
		\Require $\mathcal X_{\mathcal I},\mathcal X_{\mathcal U}$, $\mathfrak{D}_d(\cdot)$, \emph{causal} system~\eqref{eq:semi-explicit-descriptor}, horizon $T_{\max}$
        \State Select $\mathbf{z}(0)$ in $\mathcal{S}$ and impulse-free $\mathbf{u}(0)$
        \State Record $\mathbb{U}_0,\;\mathbb{Z}_0,\;\mathbb{D}_0,\;\mathbb{Z}_1$ for $T=0$
		\For{$T = 1, \ldots, T_{\max}$} 
        \State Select safe control input $\mathbf{u}(T)$
			\State Extend $\mathbb{U}_0,\;\mathbb{Z}_0,\;\mathbb{D}_0,\;\mathbb{Z}_1$
			\If{$\mathrm{rank}(\mathbb{D}_0) < N_d$}
				\textbf{continue}
    			\EndIf \textcolor{blue}{\Comment{Replace}}
			\State Solve~\eqref{eq:ct_CBC_data1} and~\eqref{eq:ct_CBC_data3} for $P$ and $\beta_1$ \textcolor{blue}{\Comment{\eqref{eq:corollary-conditions}$\rightarrow$\eqref{eq:ct_CBC_data3}}}
			\If{infeasible} \textbf{continue}
			\EndIf
			\State Optimize for $\beta_2 = \min_{\mathbf{z}\in\mathcal{X}_{\mathcal{U}}} \mathfrak{B}(\mathbf{z})$ \If{$\beta_2 \leq \beta_1$} \textbf{continue}
			\EndIf
			\State \textit{Autonomous Takeover! $\mathbf{u}$ is safe for all $k > T$.}
            \State \textbf{return} safe controller $\mathbf{u} = \mathbb{U}_0\mathbb{D}_{0}^\dagger\mathfrak{D}_d(\mathbf{z})$ certified by CBC $\mathfrak{B}(\mathbf{z}) = \mathbf{z}^\top P\mathbf{z}$ \textcolor{blue}{\Comment{Controller $\mathbf{u} = \mathbb{U}_0H(\mathbf{z})P\mathbf{z}$}}
		\EndFor
		\State \textbf{return} infeasible (no CBC safety certificate found)
	\end{algorithmic}
	\caption{Autonomous takeover by \emph{safety controller} via CBC using Theorem~\ref{thm:data-CBC} \textcolor{blue}{(Corollary~\ref{cor:CLF})}}
	\label{alg:descriptor}
\end{algorithm}

The algorithm is run as an iterative loop (Step $3$), initialized with the initial state and input at $T=0$ (Step $1-2$). In Step $4$ the user must choose a safe control input, since the user is manually controlling the system until safe autonomous takeover is possible. In Step $5$ we collect the data matrices, appending new collected data to the matrices acquired at $T-1$.  In Step $6-7$ we check if the data is persistently excited, if not we \emph{continue} (skip the remaining steps and move onto next loop iteration with $T=T+1$). Until at least $T=N_d$ the full-row rank condition will not be satisfied, and so the user will need to be controlling the system manually. In Step $8-10$ we solve for $P$ and $\beta_1$, (\emph{e.g.} using semidefinite programming), if the solution is infeasible we \emph{continue}. For the alternate controller design, we can solve for $P$, $H(\mathbf{z})$ and $\beta_1$ by replacing condition~\eqref{eq:ct_CBC_data3} with~\eqref{eq:corollary-conditions}. In Step $11-13$ we optimize for the level set $\beta_2$, and if $\beta_2 \leq \beta_1$ we \emph{continue}. Note Steps $8-13$ require separate steps even when solvable as LMIs due to the difficulty of solving for $P$ and $P^{-1}$ simultaneously. By solving $\beta_1$ with $P^{-1}$ there is a higher likelihood of finding a valid CBC compared with previous hierarchical methods~\cite{samari2024singletrajectory}. Reaching Step $14$ guarantees the design of a safety controller for both approaches, where the controllers have slightly different formulations. Reaching this stage allows the user to fully transfer control to the autonomous policy with safety guaranteed for all future time steps. The safe controller and CBC are returned in Step $15$. The algorithm terminates if no safe controller was designed within $T_{\max}$ iterations in Step $17$. $T_{\max}$ can be selected when the number of decision variables becomes too large, and the algorithm could be reinitialized at any point since the user maintains safe system control. Similarly, a sliding window, could be considered to reduce the computational complexity of finding a safe control policy as long as the PE condition is satisfied.

Each iteration of Algorithm~\ref{alg:descriptor} requires a rank test on $\mathbb{D}_0\in\reals^{N_d\times T}$ (Step $6$), costing $\mathcal{O}(N_d^2T)$ operations, followed (once PE holds) by an SDP whose cost scales with the \emph{projected} dimension $n_d$ rather than the full state dimension $n$. The data horizon $T$ enters through the pseudoinverse $\mathbb{D}_0^\dagger$ ($\mathcal{O}(N_d^2T)$) and, for the alternate design of Corollary~\ref{cor:CLF}, through the decision variable $H(\mathbf{z})\in\reals^{T\times n_d}$, motivating the sliding window above for larger systems.

In the same way that the initial region and unsafe region can be updated online when searching for a safe control policy and CBC. It is also possible to expand the extensive dictionary to consider additional nonlinear terms online. By storing $\mathbb{Z}_0$ one can add additional nonlinearities and use $\mathbb{Z}_0$ to build a broader (more extensive dictionary) $\mathbb{D}_0$. Finally, further flexibility in the controller design comes from the decomposition of $\mathfrak{D}_d(\mathbf{z})$, since $\mathfrak{D}_d(\mathbf{z}) = \Theta_d(\mathbf{z})\mathbf{z}$ does not have a unique $\Theta_d(\mathbf{z})$. Overall, the control approach is highly flexible and the conditions to satisfy are relatively lightweight.

\section{Simulation Results}
\label{sec:case-studies}
We present two unstable nonlinear case studies, the latter with a trigonometric nonlinear term. We do not consider spurious nonlinearities although they are supported by the theory, \emph{e.g.} see~\cite{samari2024singletrajectory}. Both experiments complete online synthesis in under $0.5\,$s on an Intel Core Ultra 9 285K processor with 64GB RAM.

\subsection{Predator-Prey Ecological Model}

\begin{figure}[!t]
    \centering
    \includegraphics[width=0.97\linewidth]{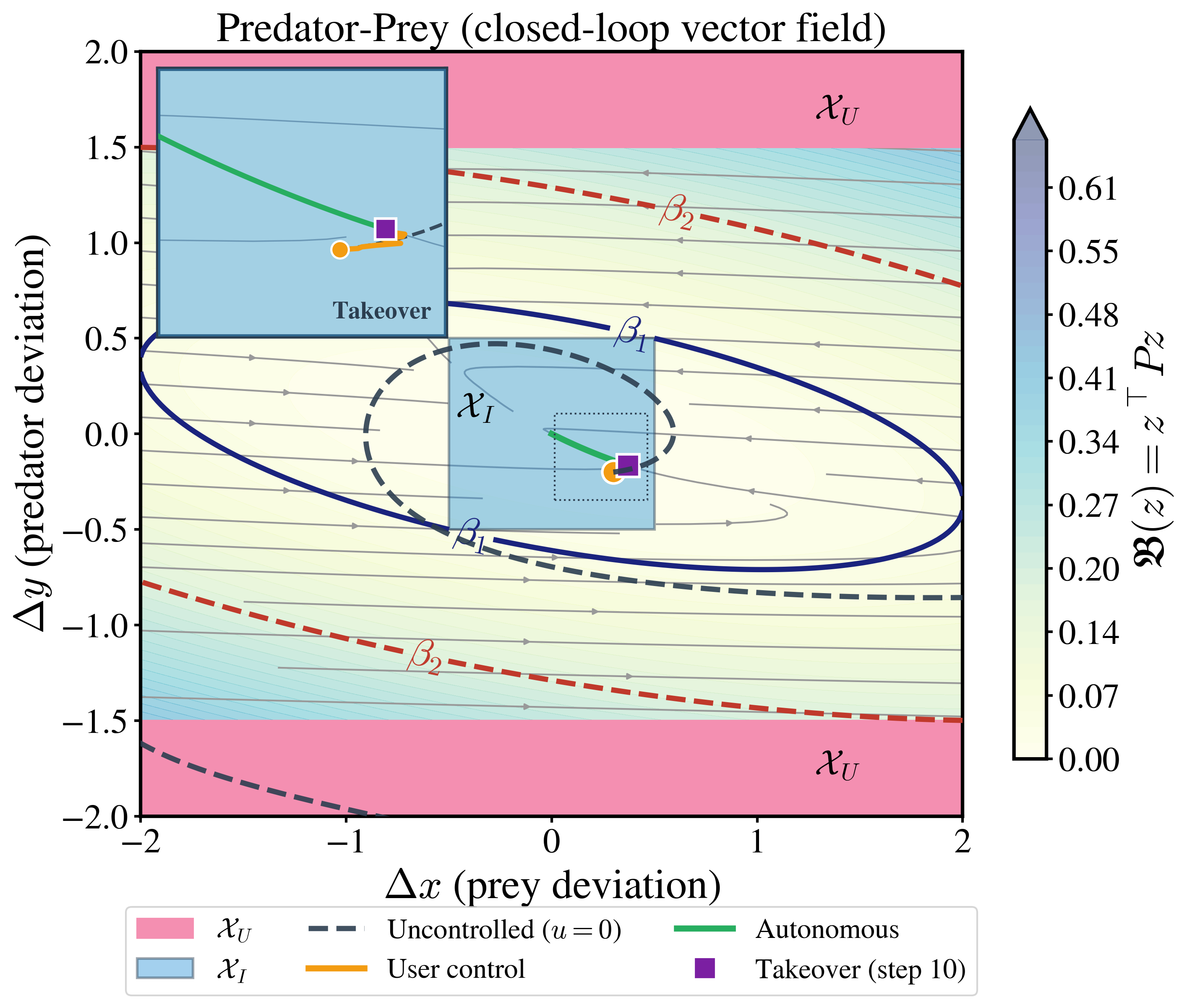}
    \caption{Demonstration of safe autonomous takeover, the vector field shows the closed-loop system with the autonomous controller applied. Takeover occurs at $k=10$ (yellow changes to green), with the uncontrolled trajectory (gray-dashed) causing ecological crisis by reaching the unsafe region $\mathcal{X}_{\mathcal{U}}$ (bottom left corner). A zoom of the dotted area of the figure is displayed in the top left corner.}
    \label{fig:pred_prey}
\end{figure}

We consider the following variation on the famous Lotka-Volterra predator-prey ecological system:
\begin{subequations}
\begin{align}
\label{eq:pred-prey}
    x(k+1) &= x(k) -h(br_1y(k) - bx(k)y(k) - u(k)),\\
    y(k+1) &= y(k) + h(dr_2x(k) + dx(k)y(k) + q(k))\\
    \label{eq:pred-prey3}
    0 &= ex(k) + fy(k) + q(k),
\end{align}
\end{subequations}
where $x(k)$ and $y(k)$ are the relative prey and predator populations, respectively, and $q(k)$ is a harvesting quota, introducing an algebraic constraint. This quota introduces greedy predators, when the predators increase relative to the prey then the predators become competitive and cull even more prey. This amplifies each cycle leading to a potential ecological crisis if the predator population deviates from the prey more than $\pm1.5$ units. The constant terms are; prey birth rate $a=1.0$, predator birth rate $b=0.5$, predator death rate $c=0.8$, predation benefit $d=0.3$, harvesting sensitivity to prey $e=0.2$ and harvesting sensitivity to predators $f=-0.3$, $h=0.05$ is the discretization parameter, and $r_1 = \frac{c}{d}$ and $r_2 = \frac{a}{b}$. The input $u(k)$ is the conservationists intervention on the prey population.

Converting this system into the descriptor form gives $E = \mathrm{diag}(1,1,0)$ which is singular. We remind the reader that we consider all coefficients to be unknown in our experiments but assume we know the dictionary $\mathfrak{D}_d(\mathbf{x}_d) = \Theta_d(\mathbf{x}_d)\mathbf{x}_d = [x,~y,~xy]^\top$ where $\mathbf{x}_d = [x,~y]^\top$ and $\mathbf{x}_a = q$. We consider an initial region $\mathcal{X}_\mathcal{I} = [-0.5,0.5]^2$ and unsafe regions $\mathcal{X}_\mathcal{U} = \{|y| \geq 1.5\}$ (describing ecological crisis). 

We collect data according to Assumption~\ref{ass:subspace-data}, where we consider 
$\mathbf{z}^+ = \Tilde{A}\mathfrak{D}_d(\mathbf{z}) + \Tilde B\mathbf{u}$ to be the system~\eqref{eq:pred-prey} projected onto its subspace $\mathcal{S}$~\eqref{eq:pred-prey3}, where $\mathbf{z} = \mathbf{x}_d$. Figure~\ref{fig:pred_prey} demonstrates the results of running the case study. The system is initialized at $\mathbf{z}(0)=[0.3, -0.2]^\top$. The user then selects the following safe inputs (yellow line): 
\begin{align*}\mathbb{U}_0 = [ 0.286, -0.184, 0.163, 0.474, -0.608,\\ -0.562, 0.383, -0.204, -0.869, -0.253 ],\end{align*} 
before the system declares it can complete \emph{autonomous takeover} when $k=10$ at $\mathbf{z}(10)=[0.372,-0.165]^\top$ marked by the purple box. Based on the user input, the full-row rank condition was satisfied at $k=10$. The predator-prey system is polynomial so we leverage the well-established S-procedure in step $8$ of the algorithm and solve for a feasible $P$ and $\beta_1$ via SDP. $\beta_2$ is computed simply in step $11$ using convex optimization over the semi-algebraic region $\mathcal{X}_{\mathcal{U}}$. The subsequent autonomous controller (green line) is given by, 
    \[\vspace{-0.15cm} u(k) =-0.608x(k) - 15.508y(k) + 41.570x(k)y(k)\vspace{0.01cm}\]
and certified by the CBC, \[\mathfrak{B}(\mathbf{z}(k)) =0.014x(k)^2 + 0.0406x(k)y(k) + 0.1119y(k)^2,\] with level sets $\beta_1 = 0.0416$ and $\beta_2 = 0.1858$. Without human intervention, and the safe autonomous takeover, the system would fall into ecological crisis, shown by the gray-dashed trajectory reaching the unsafe region (bottom left corner).

\begin{figure}[!t]
    \centering
    \includegraphics[width=0.97\columnwidth]{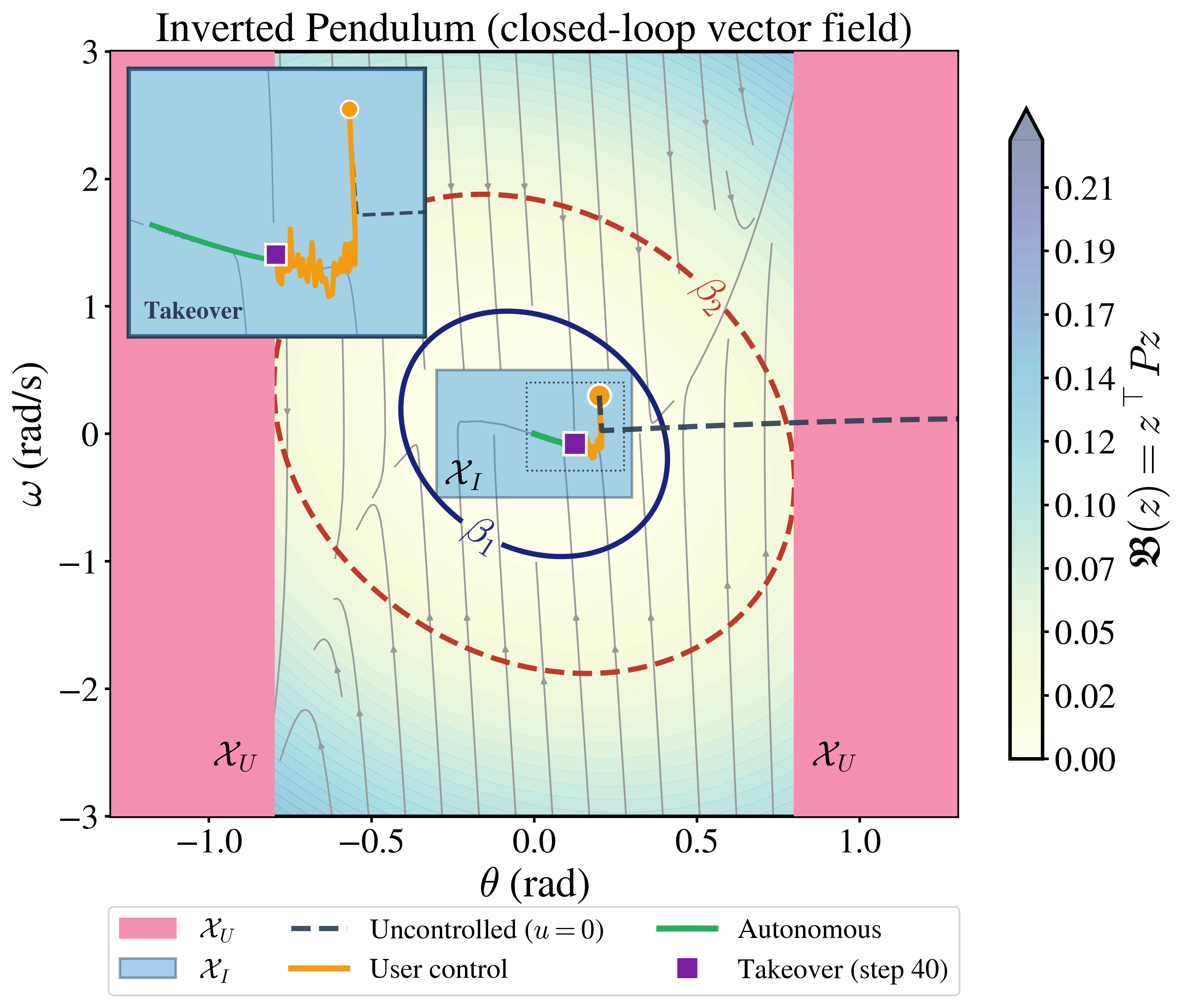}
    \caption{Demonstration of safe autonomous takeover for the inverted pendulum. The vector field shows the closed-loop system with autonomous controller applied. Takeover occurs at $k=40$ (yellow changes to green), with the uncontrolled trajectory (gray-dashed) causing the pendulum to fall by reaching the unsafe region $\mathcal{X}_{\mathcal{U}}$. A zoom of the dotted area of the figure is displayed in the top left corner.}
    \label{fig:inv_pend}
\end{figure}

\subsection{Inverted Pendulum}

As our second example, consider an inverted pendulum mounted on a fixed pivot, coupled to a rotary viscous damper whose reaction torque $\tau_r$ enters as an algebraic variable:
\begin{subequations}
\label{eq:inv-pend}
\begin{align}
    \theta(k+1) &= \theta(k) + h\omega(k),\\
    J\omega(k+1) &= (J-hb)\omega(k) + hmgl\sin(\theta(k)) \nonumber\\&\quad+ h\tau_r(k) + hu(k),\\
    \label{eq:inv-pend3}
    0 &= c\omega(k) + \tau_r(k),
\end{align}
\end{subequations}
where $\theta(k)$ is the angle from the upright equilibrium and $\omega(k)$ is the angular velocity. The algebraic constraint~\eqref{eq:inv-pend3} models a viscous rotary damper in equilibrium with $\tau_r(k) = -c\omega(k)$. The constant terms are; pendulum mass $m=1.0$kg, gravitational acceleration $g=9.81$m/s$^2$, pendulum length $l=0.5$m, rotational inertia $J=0.25$kgm$^2$, pivot friction $b=0.5$Nms/rad, and discretization parameter $h=0.02$. The input $u(k)$ is a control torque applied at the pivot. Without active control, the gravitational torque drives the pendulum to the unsafe region (gray-dashed line in Figure~\ref{fig:inv_pend}). We consider an initial region $\mathcal{X}_\mathcal{I} = [-0.3,0.3]\times[-0.5,0.5]$ and unsafe region $\mathcal{X}_\mathcal{U} = \{|\theta| \geq 0.8\}$ (pendulum too far from vertical). Similar to the previous case, the descriptor form here gives $E = \mathrm{diag}(1, J, 0)$ which is singular. We consider the dictionary $\mathfrak{D}_d(\mathbf{x}_d) = [\theta,~\omega,~\sin(\theta)]^\top$ where $\mathbf{x}_d = [\theta,~\omega]^\top$ and $\mathbf{x}_a = \tau_r$.

The inverted pendulum dictionary contains the nonconvex transcendental function $\sin(\theta)$, which prevents a direct SDP formulation, but it can be factorized as $\sin(\theta) = \mathrm{sinc}(\theta) \theta$. As $\mathrm{sinc}(\theta)$ is concave over the domain, if the LMI evaluated
at both end-points is non-negative then the interval is nonnegative everywhere; $f(s) \geq 0$
on $[a, b]$ if and only if $f(a) \geq 0$ and $f(b) \geq 0$. This statement
is trivial but underutilized for CBCs. Thus $s = \mathrm{sinc}(\theta) \in [s_{\min},1]$ with $s_{\min} = \sin(0.3)/0.3 \approx 0.985$. Substituting $F=\mathbb{Z}_1\mathbb{D}_0^\dagger$ and $Q=P^{-1}$ in~\eqref{eq:ct_CBC_data3} gives an LMI in $Q$, where every entry is linear in $Q$ for fixed $\mathbf{z}$. Exploiting the structure of $\Theta_d(\mathbf{z})$, and the vector $F\Theta_d(\mathbf{z}) = C_0 + sC_1$ where $s = \mathrm{sinc}(\theta)$:
\begin{equation*}
    \begin{bmatrix}
        Q & (C_0 + sC_1)Q\\
        Q(C_0 + sC_1)^\top & Q
    \end{bmatrix} \succeq 0, \quad \forall s\in[s_{\min}, 1].
\end{equation*}
No Taylor approximation is required.  The guarantees are robust and the Lyapunov decrease~\eqref{eq:ct_CBC_data3} holds for all $\mathbf{z}\in\mathcal{X}_\mathcal{I}$. We calculate $\beta_1$ and $\beta_2$ using the same approach as for predator-prey.

Collecting the data according to Assumption~\ref{ass:subspace-data}, we consider $\mathbf{z}^+ = \tilde{A}\mathfrak{D}_d(\mathbf{z}) + \tilde{B}\mathbf{u}$ to be the system~\eqref{eq:inv-pend} projected onto its subspace $\mathcal{S}$~\eqref{eq:inv-pend3}, where $\mathbf{z} = \mathbf{x}_d$. Figure~\ref{fig:inv_pend} demonstrates the results of running the case study. The system is initialized at $\mathbf{z}(0) = [0.2, 0.3]^\top$. The user then selects the following safe inputs (yellow line):

{\small\begin{align*}
\mathbb{U}_0 = [&-4.744,~ -5.950,~ -4.306,~ -2.510,~ -6.005,~ -5.934, \\
&-2.238,~ -3.847,~ -6.280,~ -4.181,~ -6.149,~ -6.081, \\
&-4.594,~ -8.855,~ -8.365,~ -5.937,~ -6.770,~ -4.034, \\
&-6.436,~ -7.366,~ -1.519,~ -4.896,~ -4.251,~ -7.188, \\
&-5.336,~ -3.964,~ -6.445,~ -3.311,~ -5.230,~ -4.549, \\
&-5.117,~ -0.147,~ -3.890,~ -5.932,~ -2.097,~ -6.167, \\
&-3.229,~ -7.533,~ -6.172,~ -3.046],
\end{align*}}

\noindent before the system declares it can complete \emph{autonomous takeover} when $k=40$ at $\mathbf{z}(40) = [0.126, -0.077]^\top$ marked by the purple box. The PE condition was satisfied at $k=10$, but Step $8$ of Algorithm~\ref{alg:descriptor} remained infeasible until $k=40$ as additional data was required to sufficiently capture the nonlinear dynamics. The subsequent safe autonomous controller (green line) is given by,
\[
\vspace{-0.2cm}
    u(k) = 1919.1\,\theta(k) - 0.031\,\omega(k) - 1958.4\,\sin(\theta(k)),
    \vspace{0.01cm}
\]
and certified by the CBC,
\[
\vspace{-0.2cm}
    \mathfrak{B}(\mathbf{z}) = 0.0581\,\theta(k)^2 + 0.0100\,\theta(k)\omega(k) + 0.0105\,\omega(k)^2,
\]
with level sets $\beta_1 = 0.0094$ and $\beta_2 = 0.0356$. Without human intervention and the safe autonomous takeover, the uncontrolled system diverges into the unsafe region within seconds as the pendulum falls over (grey-dashed trajectory in Figure~\ref{fig:inv_pend}).

\vspace{-0.2cm}
\section{Conclusion}
\label{sec:conclusion}

In this work we provide the first design of a formal safety controller for an unknown discrete-time nonlinear descriptor system certified by a data-driven control barrier certificate (CBC). We present an online algorithmic approach, including two controller designs, enabling \emph{autonomous takeover} (safe transfer of control) from a user to an autonomous control policy guaranteeing safety. We demonstrate our approach on unstable descriptor systems, including an inverted pendulum with a trigonometric nonlinear term (without needing approximation methods).

\bibliographystyle{IEEEtran}
\bibliography{Bibliography}

\end{document}